\documentclass[aps,pre,onecolumn,floatfix,nofootinbib,amsmath,amssymb]{revtex4}
\usepackage{graphicx}
\usepackage[english]{babel}
\usepackage{graphicx}
\usepackage{textcomp}
\usepackage{amsmath}
\usepackage[dvips]{color}
\usepackage{dcolumn}
\usepackage{bm}
\begin{document}

\title{Universality and scaling in human and social systems}

\author{Chin-Kun Hu$^{1,2}$}
\email{huck@phys.sinica.edu.tw}

\affiliation{$^1$Institute of Physics, Academia Sinica, Taipei 11529, Taiwan}

\affiliation{$^2$Department of Physics, National Dong Hwa University, Hualien 97401, Taiwan}


\begin{abstract}
The objective of statistical physics is to understand macroscopic behavior of a many-body system from the
interactions of the constituents of that system. When many-body systems reach critical states, simple universal
 and scaling behaviors appear. In this talk, I first introduce the concepts of universality and scaling
  in critical physical systems, I then briefly review some examples of universal and scaling behaviors
  in human and social systems, e.g. universal crossover behavior of stock returns,
  universality and scaling in the statistical data of literary works,
   universal trend in the evolution of states or countries etc.
   Finally, I mention some interesting problems for further studies.
\end{abstract}

\maketitle

\section{Introduction}




"Universality" means applicable in different content; "scaling" means physical quantities from different system sizes
as functions of the parameter, e.g. temperature, of the systems can
collapse approximately on a single curve by a simple change of scale for each quantity \cite{05lit}.
A good example of scaling is Russian doll, which contains a set of self-similar dolls of different length scales.
One of the objectives of scientific research is to use a small number of concepts, or functions, or laws to describe a large numbers
of natural or laboratory phenomena. Thus it is not surprising that the concepts of universality and scaling
have been used by researchers in different branches of sciences to describe behavior of the studied systems \cite{05lit}.

Universality is the key idea in the search for fundamental interactions of matter.
Issac Newton (1642-1727 A.D.) discovered that the falling of apples to the earth, the motion of the moon
around the earth, and the motion of the earth around the sun are governed by the same physical laws: universal
 law for gravitational forces of matter and Newton's laws of motion. James Clerk Maxwell (1831-1879 A.D.)
  proposed Maxwell's equations to describe electric and magnetic interactions. In recent decades,
  S. Weiberg and A. Salam proposed a unified gauge field theory to describe the weak and electromagnetic interactions.
  The major objective of theoretical elementary particle physics is to formulate a unified theory which
  can describe electromagnetic, weak, strong, and gravitational interactions \cite{05lit}.

  Besides basic interactions, the idea of universality is also useful for the understanding of many-body interacting systems,
   especially the critical systems \cite{71Stanley,2014cjpHuCK}.
   The quantitative and systematic researches on  critical physical systems (e.g.~$\rm CO_2$ and $\rm H_2 O$) started in the later half of the 19th.
   century in the study of liquid-gas critical systems, and made tremendous progress in the 20th. century
   via the ideas and methods from statistical physics \cite{71Stanley,2014cjpHuCK}.

 In 1869, Thomas Andrews (9 December 1813 - 26 November
1885) presented a detailed report on the critical phenomena of
$\rm CO_2$ \cite{1869AndrewsT}. The data in \cite{1869AndrewsT} were used by S. G. Brush to plot isothermal curves of $\rm CO_2$
on the pressure $P$ vs. colume $V$ plane  \cite{76Brush}, which is reproduced in Fig. 1(a).

\begin{figure}
\centerline{\includegraphics[width=16 cm]{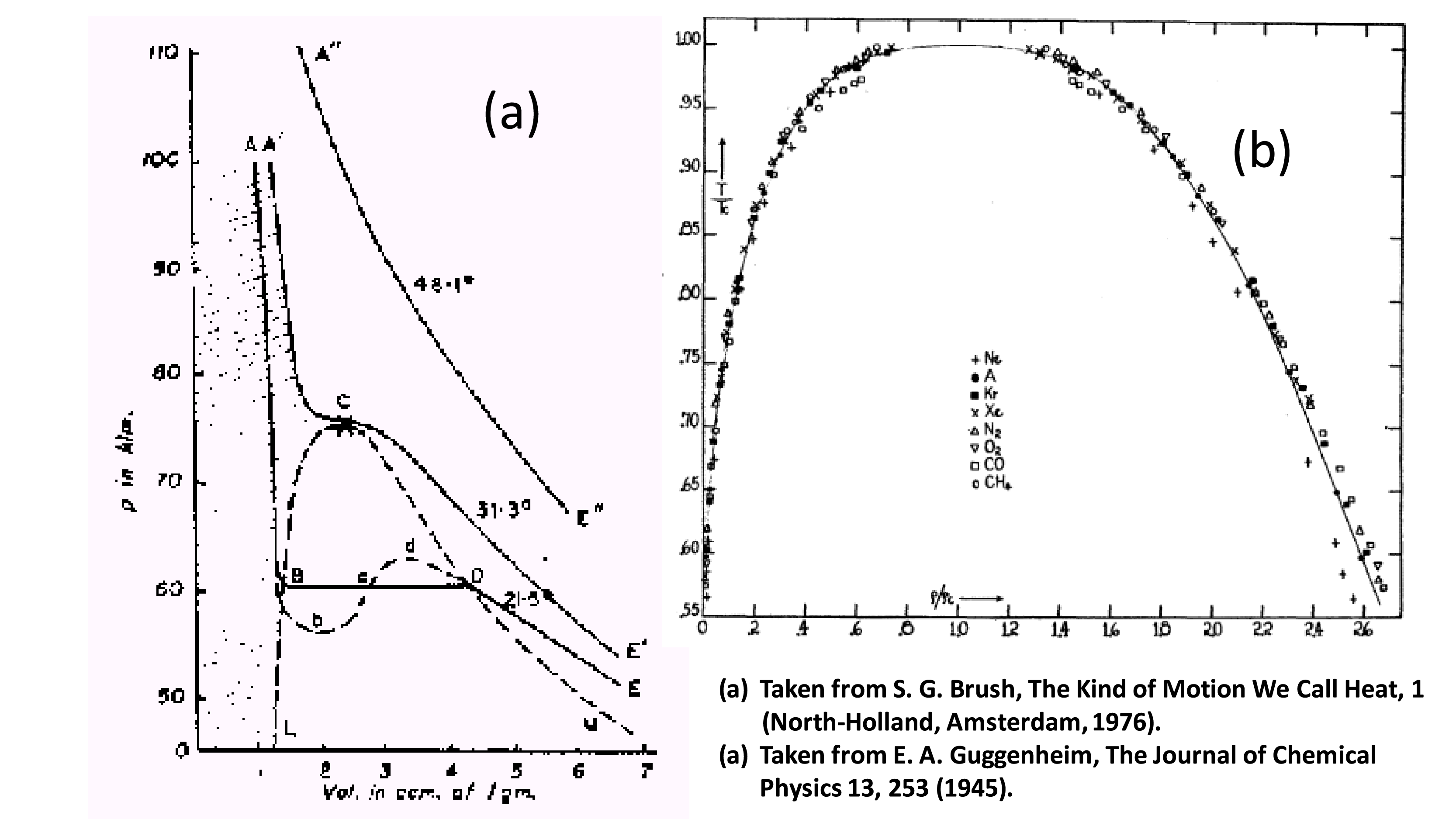}}
\hfill \caption{(a) Isothermal curves for $\rm CO_2$ in the pressure ($P$) vs. volume ($V$) plane plotted by S. G. Brush \cite{76Brush} who used
the data reported in  \cite{1869AndrewsT} by Thomas Andrews in 1869. (b) Collapse of data for 8 gas-liquid systems on the universal curve
with a critical exponent 1/3. Taken from  \cite{45jcp} by E. A. Guggenheim.} \label{CPFig1}
\end{figure}

In 1834, Emile Clapeyron (1799-1864) proposed the ideal gas law $PV=RT$
for a mole ($\approx 6.02 \times 10^{23}$ molecules or atoms) of ideal gas,
where $R$ is the ideal gas constant, and $T$ is the absolute
temperature.
In 1873 Johannes Diderik van der Waals (23 November 1837 - 8
March 1923)  took into account finite sizes and interactions of
gas molecules to modify the ideal gas law and proposed that a mole
 of liquid-gas system can be described by the
equation \cite{71Stanley,2014cjpHuCK,1873vanderWaals,86HuCK}
\begin{equation}
\left( P+\frac{a}{V^2} \right)(V-b)=RT, \label{vdw}
\end{equation}
where $a$ and $b$ are constants. One can use Eq.~(\ref{vdw}) to plot $P$ as a function
of $V$ for a fixed $T$ in the $P$-$V$ plane. For a constant $T <
T_c$ and a constant $P$, Eq.~(\ref{vdw}) could have three
solutions for $V$; in this case Maxwell construction (an equal
area rule for the parts below and above horizontal line) can be
used to obtain a curve in the $P$-$V$ plane, which has a
horizontal portion corresponding to liquid-gas coexistence region
\cite{71Stanley}. When $T$ is increased, the size of this region
decreases. When $T$ reaches the critical temperature $T_c$, the
size of this region becomes 0. From this condition, one can derive
that \cite{71Stanley,2014cjpHuCK}
$V_c=3b, \quad P_c={a}/{27b^2}, \quad T_c={8a}/{27bR}$,
from which one can determine $a$, $b$, and ideal gas constant $R$
from measured values of $V_c$, $P_c$, and $T_c$.
Two curves which connect boundaries of the horizontal portions for
$T \le T_c$ form phase boundaries for liquid phase,
liquid-gas coexistence phase, and gas phase. For  $T \gg T_c$,
Equation (\ref{vdw}) can be approximated by the ideal gas law
$PV=RT$.


In 1881, van der Waals defined reduced volume $\phi$, reduced pressure $\pi$, and reduced
temperature $\theta$ as
$\phi=V/V_c$, $\pi=P/P_c$, $\theta=T/T_c$,
then he rewrote Eq.~(\ref{vdw}) as
\begin{equation}
\left( \pi+\frac{3}{\phi^2} \right)(3\phi-1)=8\theta, \label{rvdw}
\end{equation}
which is called reduced van der Waals' equation and means that in terms of reduced
 variables, data for different substance
can collapse on the same curve.
In 1910, van der Vaals received the Nobel Prize in physics
``for his work on the equation of state for gases and liquids''.

In 1924-1931,  Lennard-Jones \cite{24LJ,31LJ} proposed that
the interaction potential between a pair of atoms or molecules at
sites $i$ and $j$ with the distance $r_{ij}$ is
\begin{equation}
v_{ij}(r_{ij})=4 \epsilon \left(
\left( \frac{\sigma}{r_{ij}} \right)^{12} -\left( \frac{\sigma}{r_{ij}} \right)^{6} \right),
\label{L-J}
\end{equation}
where $\sigma$ and $\epsilon$ are used as units for length and
energy strength, respectively. Thereafter, the potential of Eq.~(\ref{L-J}) is called Lennard-Jones potential.

 In 1945, Guggenheim \cite{45jcp}
reported that in the $T/T_c$ versus ${\rho}/\rho_c$ plane
($\rho$ and $\rho_c$ are density and critical density of the
substance, respectively)  the coexistence curves of ${\rm N_e}$,
${\rm A_r}$, ${\rm K_r}$, ${\rm X_e}$, ${\rm N_2}$,  ${\rm O_2}$,
${\rm CO}$, and ${\rm CH_4}$ are very consistent with each other,
which is reproduced in Fig. 1(b).
Thus van der Waals's general idea for the universality of the
coexistence curves was confirmed. However, near the
critical point, Guggenheim  found that
\begin{equation}
 \frac{\rho_l -\rho_g }{\rho_c} =\frac{7}{2}  |1-\frac{T}{T_c}|^{\beta},
 \label{45cjp}
 \end{equation}
where the critical exponent $\beta$ is 1/3 \cite{45jcp}, on
the other hand van der Waals's equation gives $\beta =1/2$ \cite{71Stanley}. The
discovery of such deviation marks
modern era of critical phenomena \cite{71Stanley}.

In1952, Yang and Lee proposed that the critical behavior of the
 gas-liquid system can be represented by a lattice-gas model
\cite{YangLee52}, which is equivalent to the Ising model \cite{25Ising},
in which each atom or molecule on a lattice site is assigned a variable which
can be either +1 or -1.




In 1960s, many experimental and theoretical results indicate the universality and scaling in
critical physical systems \cite{71Stanley,2014cjpHuCK}. It has been proposed that critical systems
can be classified into different universality classes so that the systems in the same class has the
same set of critical exponents. It has been found that critical exponents satisfy certain equalities,
call scaling laws or relations \cite{71Stanley,2014cjpHuCK}.
The phenomenal theory had been proposed  to understand the scaling relations \cite{71Stanley,2014cjpHuCK}.
In 1971-1972,  Ken G. Wilson developed renormalization group (RG)
theory \cite{71prbRGa,72prlRGa}  for critical phenomena,
which provides a framework for understanding  universality and scaling in
critical systems, and can be used to calculate critical quantities of
critical systems. For this contribution,  Ken G. Wilson won the Nobel Prize in physics
in 1982 \cite{83rmp-WilsonKG}.

%


\begin{figure}
\centerline{\includegraphics[width=16 cm]{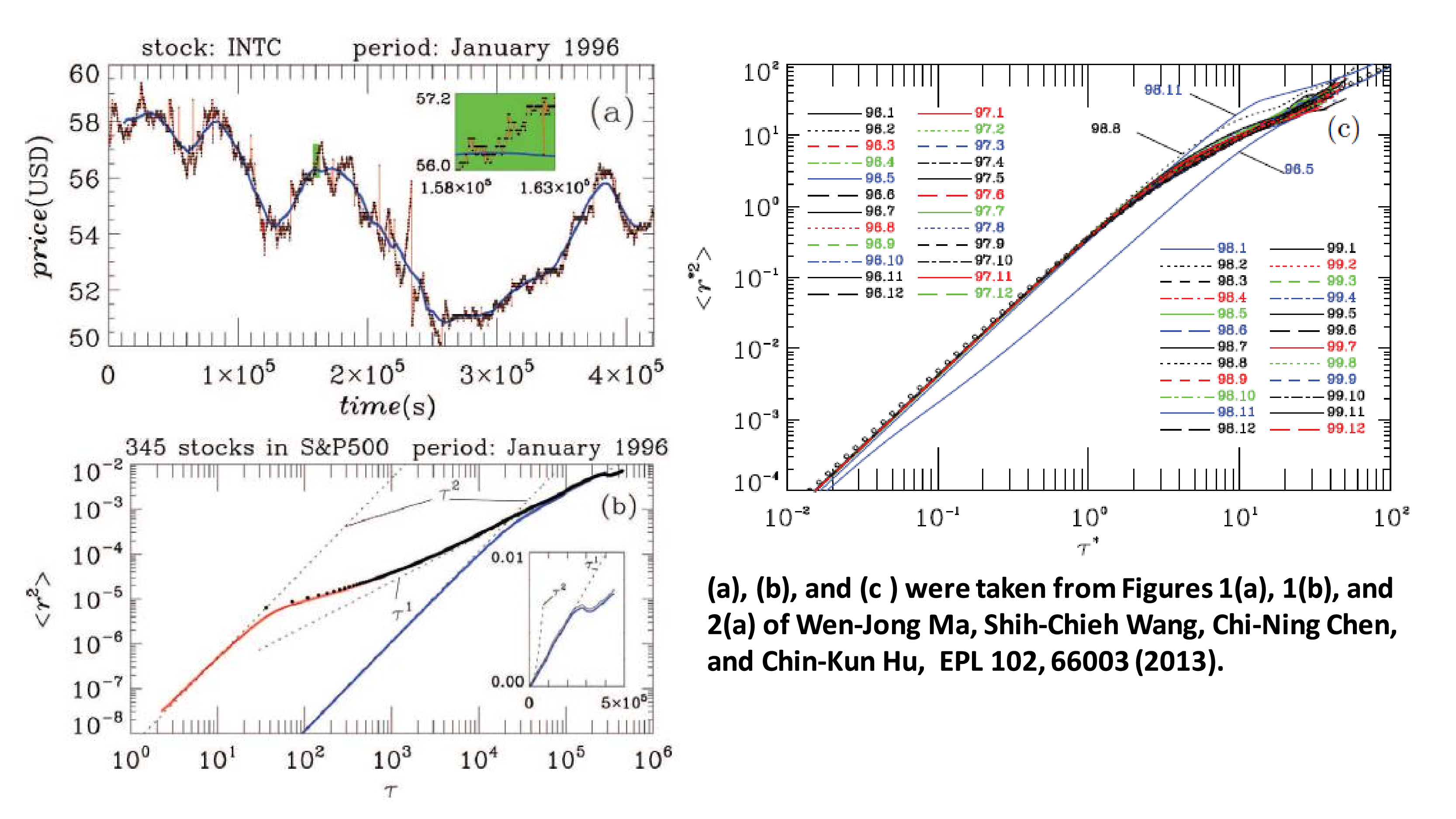}}
\hfill \caption{(Color online)(a) Stock prices of Intel Corporation
(INTC), over 18 trading days in January 1996. The market data,
collected in 36-seconds intervals, are marked by dots, which are
connected by red lines and their one-day moving averages in
steps of 36-seconds are plotted by the blue line. The colored
inset shows the enlarged image of the region marked by the same
(green) color. (b) Log-log plot for the mean-square-log-returns
(MSLR) versus time interval $\tau$, for a collection of 345 stocks
as in \cite{MHA}, including INTC, picked from S\&P500 over January 1996,
calculated either based on original data (marked by dot points),
or based on the moving-average (blue curve) shown in (a). The
red line shows the result over short time scales, based on the
continuous (red) curve of (a); the blue curve shows the results
obtained from the blue curve in (a). In the inset, we show the
plots the two former curves in linear scales. (c) Scaled mean square log-return
$<r^{*2}>$ versus scaled time internal $\tau^*$, for each month in
1996-1999, for  the 345 stocks in S\&P500 in US that have been analyzed
in (b). They are compared
with the master curve (open circles) defined by
Eq.~(\ref{MSDmaster}).  The time sequences for the collections of data
 are in intervals of 36 seconds. The fitting to asymptotic
form $\theta t^\alpha_1$ and $D t$ are carried out over the ranges
$36<\tau<7200$ and $54000<\tau<108000$. Those months with their scaled MSLRs significantly
deviating from the master curve are marked specifically. } \label{StockFig1}
\end{figure}

In 1995-1996, Bl$\ddot{o}$te and collaborators
\cite{95jpa3dIsing,96jpaIsing} used Monte Carlo simulations to
find that the critical exponent $\beta$ of the spontaneous
magnetization and $\nu$ of the correlation length \cite{71Stanley}
of a three-dimensional (3D) Ising model are 0.3269(6)
\cite{96jpaIsing} and 0.6301(8) \cite{95jpa3dIsing}, respectively.
In 2009, Sengers and Shanks \cite{review} reviewed the
experimental data for liquid-gas critical systems and reported
that $\beta=0.3245$ and $\nu=0.629 \pm 0.003$, respectively.
In 2012, Watanabe, Ito and Hu \cite{2012jcp} used molecular
dynamics simulations to find that $\beta$ and $\nu$ of a 3D
Lennard-Jones (L-J) model system \cite{24LJ,31LJ} are 0.3285(7)
and 0.63(4), respectively. The values of $\beta$ and $\nu$
obtained for simple model systems reported in Refs.
\cite{95jpa3dIsing,96jpaIsing,2012jcp} are highly consistent with
experimental data reported in \cite{45jcp, review}.
Using Monte Carlo methods \cite{49MC,53jcpMC,92prbHu} and analytic methods \cite{02jpaWuMC,02jpaIIH}
it has been found that many critical systems can have very nice universal and scaling behaviors
\cite{84PF,94jpaHu,95prlHu,96prlHu,97preHu,98preLinCY,99preA,99preB,01prlIH,03preWHI}.

It is of interest to know whether the concepts and methods from
critical physical systems can be useful for understanding complicated biological and human-social systems.
In previous papers \cite{07bjLiMS,2013aipHuCK,11eplratio,11EPLSasun,13PlosOneSasun,2014HDCKHu,2015aggHuCK,15SR,17SR},
we had pointed out that the ideas and methods of statistical
physics can be useful for understanding some interesting biological problems.
In this paper, I would like to point out that such methods and concepts can also be useful for
understanding some interesting human and social systems.




\section{Universality and scaling in stock markets}


In 1828, R. Brown published a paper about ``A brief account of microscopic observations  on the particles contained in the pollen of plants
and on the general existence of active molecules" \cite{1828BM}.

 In 1900, Louis Bachelier (1870-1946) proposed a random walk model for financial market \cite{1900RWM}.
 In 1905-1908, Albert Einstein (1879-1955) published a series of papers about random walk model for
 Brownian motion  \cite{1828BM} in Germany whose English translation can be found in \cite{1905Einstein}.

 In 1908, Paul Langevin (1872-1946) proposed a stochastic equation to describe the displacement of
 Brownian particles \cite{1908LangevinL} in French whose English translation was published in \cite{97AJP}.
 Later studies indicate that the financial market does not follow the random walk model \cite{1900RWM}.
 Some results from \cite{13eplMaWJ} by Ma, Wang, Chen and Hu are mentioned below.

 Mantegna and Stanley \cite{95nature} showed that the
scaling of the probability distribution of an economic index (S \& P 500) can be described by a non-gaussian process
for data in 1984-89.
Scaling behavior is observed for time intervals from 1,000 min to 1 min.
Saakian {\it et al.} obtained exact non-Gaussian distribution of stock returns from the multifractal
random walk model \cite{11EPLstock}.

In 1999, results from two groups indicate that the evolution of stocks can not be described by the random walk model \cite{1900RWM}.
Laloux, {\it et al.} \cite{Laloux} calculated correlation matrix for 406 stocks in S \& P 500 during 1991-1996 with time
interval of one day and Plerou, {\it et al.} \cite{Plerou1,Plerou2} calculated correlation matrix for 1000
stocks in USA during 1994-1995 with time interval of 30 minutes.
Both groups found that in the eigenvalue distribution of the correlation matrix, there are some discrete distributed larger
eigenvalues above the continuous component predicted by the random walk model for stocks. In the eigenvector corresponding to the
largest eigenvalue of the correlation matrix ($\lambda_{\rm M}$), all stocks in the market move (deviate from the average value) in
the same direction. The mode corresponding to $\lambda_{\rm M}$ is called the market mode.

After the work reported in \cite{Laloux} and \cite{Plerou1,Plerou2}, one interesting question was how to modify the
 random walk model so that the revised model can well describe the eigenvalues from stock markets. For this purpose,
Ma, Hu and Amritkar \cite{MHA} proposed a model of coupled random walks for stock-stock correlations (see also \cite{Noh});
the walks are coupled via a mechanism that the displacement (price change) of each walk (stock) is activated by
the price gradients over some underlying network. They assumed that the network has two underlying structures: one for the
correlations among the stocks of the whole market and another for those within individual groups; each with a coupling parameter
controlling the degree of correlation. The model provides the
interpretation of the features displayed in the distribution of
the eigenvalues for the correlation matrix of real market on the
level of time sequences. They verified that such a model indeed
reproduces the major eigenvalue spectra of the US stocks.

The mean squares of log-return for stocks are known to have an
asymptotic $\tau^{\alpha}$ dependence on the time interval $\tau$ for
large $\tau$, with the exponent $\alpha$ close to unity. Such a
property is similar to the diffusion behavior of particles in their
mean square displacement \cite{13eplMaWJ}. It is of interest to know
whether such analogy can be extended to short $\tau$ region.
To answer this question, we study the average square log return with different time interval $\tau$.

 Figure~\ref{StockFig1}(a) shows the stock price $P(t)$ of the INTEL
Corporation as a function of time $t$, for 18 trading
days in January 1996; each day has trading time 6.5 hours=23400
seconds. For the convenience of analysis, the data are collected
in the time interval of 36 seconds. For the price changes of a
collection of stocks, we consider the analogy between the particle
displacement and the log-return $r(t,\tau)=
\log(P(t+\tau))-\log(P(t))\approx (P(t+\tau)-P(t))/P(t)$ for the
price $P(t)$ over an interval $\tau$ starting at time $t$. The
log-return carries the changes relative to the prices so that it
is a quantity that effectively renders all stocks on an equal
foot, despite of the inherited heterogeneities among the companies
and their stocks prices. The dot points in Fig.~\ref{StockFig1}(b)
shows the mean square log-return (MSLR) $<r^2>$ of a collection of
345 stocks (the same as those studied in \cite{MHA}) of the S\&P
500, versus $\tau$ during the month January of 1996. The average
$<\cdot>$ is taken over all the events for any time interval $(t,
t+\tau)$ from the 36-second-step data, for any of the 345 stocks.
The result does share the major features with the mean square
displacement (MSD) of particles, that makes possible an effective
comprehension of these empirical data.

To reveal the analogy in time evolution between the stock prices
and the particle trajectories, we analyze the paths of the
``motion" of the stocks in the one dimensional price space. We
adopt two different ways to analyze the data, either by
refining the intra-day microscopic temporal features or by
eliminating those heterogeneity via a coarse-graining procedures.
For the latter, we collect the high frequency one-day moving
averages (HF1MV) $\bar{P}(t)$ of the prices for individual stocks,
by taking simple averages over a shifting window of
one-trading-day wide (23400 seconds, or 650 intervals). To retain
the high frequency feature of the data, the window shifts in steps
of 36 seconds. The blue line in Fig.~\ref{StockFig1}(b) shows the
36-second MSLR, $<r^2>$, of January 1996, calculated from the
returns $r(t,\tau)= \log(\bar{P}(t+\tau))-\log(\bar{P}(t))$ for
the collection of 345 stocks.

 A simple model to cover both time regimes, for the motion of a
tracer particle of mass $m$ in velocity $v(t)=\frac{d R}{dt}$ in
$n_d=1$ dimension, is described by the Langevin equation \cite{1908LangevinL,97AJP}
of motion with white noise $\xi (t)$
\begin{equation}
\frac{d v}{dt} = - \frac{1}{\tau_0} v (t) + \xi (t).
\end{equation}
The balance between the friction $\frac{m}{\tau_0}v (t)$ and the
random force $m\xi (t)$ imposes a condition on the amplitude of
$\xi (t)$: $<\xi (t) \xi (t')> =\frac{2m^2 v(0)^2}{\tau_0}
\delta(t-t')$. The MSD is given by \cite{StatPhys}
\begin{equation}
{\rm MSD}\equiv <(R(t+\tau)-R(t))^2>= D[\tau-\tau_0
(1-e^{-\tau/\tau_0})] \label{MSDLaggevin}
\end{equation} where
$D=2\tau_0 v(0)^2$. For the smaller and the larger $\tau$, we have
asymptotic ${\rm MSD} \approx \frac{D}{2\tau_0} \tau^2$, for $\tau
\approx 0$, and ${\rm MSD} \approx D \tau$, for $\tau >> 1$,
respectively. Note that the ratio $\mu$ between the pre-factor $D$
of the latter asymptotic expression to that $\frac{D}{2\tau_0}$ of
the former one measures the ``mobility'' of the tracer particle.
The result $\mu = 2 \tau_0$ is
Einstein's relation.

\begin{figure}
\centerline{\includegraphics[width=16 cm]{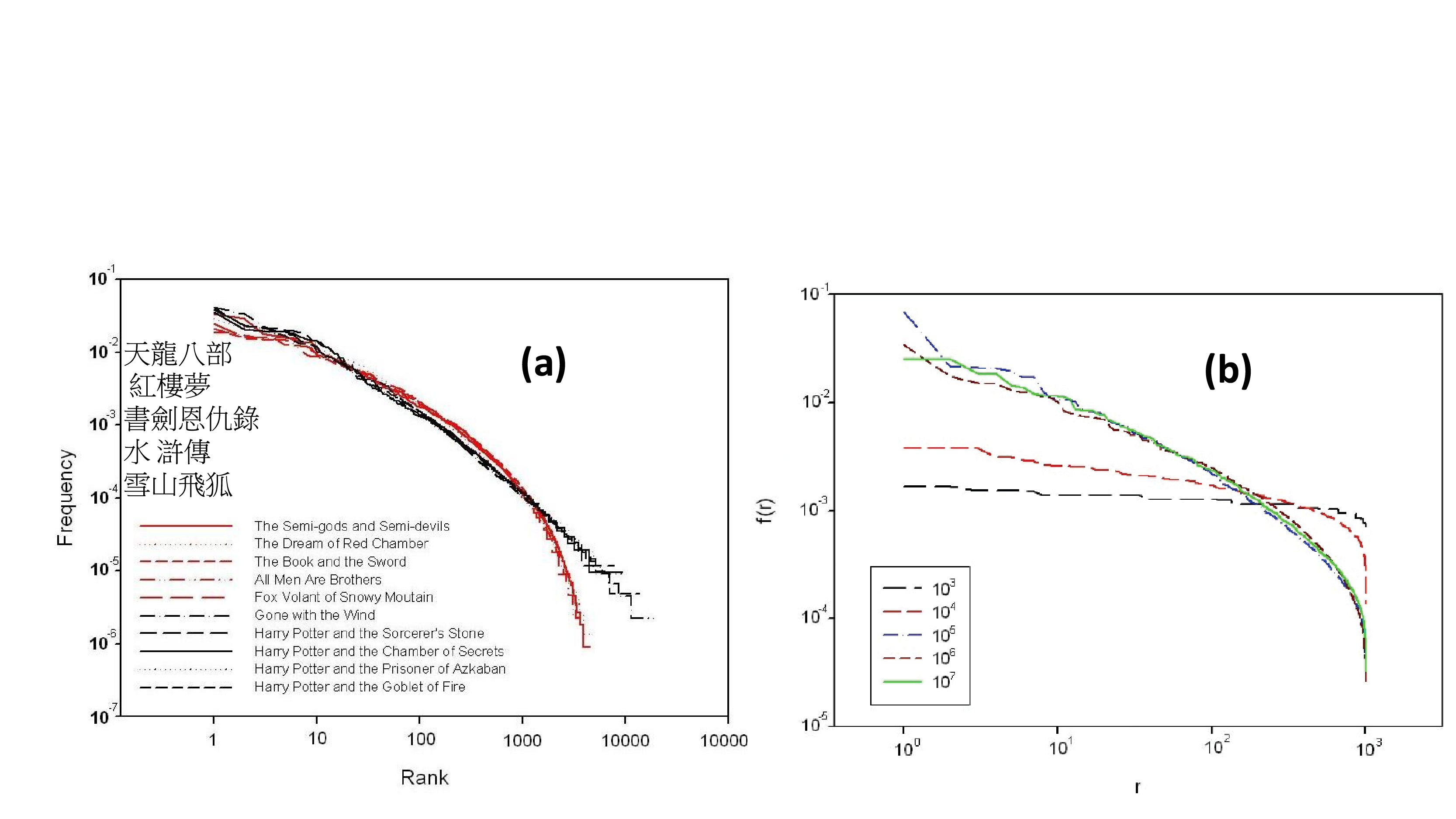}}
\hfill \caption{(Color online) (a) Zipf's distributions for five Chinese novels (The Semi-gods and Semi-devils,
The Dream of Red Chamber, The Book and the Sword, All Men Are Brothers, and Fox Volant of Snowy Moutain)
and five English novels (Gone with the Wind, Harry Potter and the Sorcerer's Stone, Harry Potter and the Chamber of Secrets,
Harry Potter and the Prisoner of Azkaban, and Harry Potter and the Goblet of Fire). The Chinese titles of Chinese novels
are listed in the same order above the English titles. Please note
that data of the former and data of the latter fall on separate curves.
(b) An example of computer simulations to generate the Zipf's distribution with time steps $10^3$,
$10^4$, $10^5$, $10^6$ and $10^7$. As time steps increase, the distribution approaches a steady distribution.
(a) and (b) were taken from Fig. 8 and Fig. 9 of \cite{05lit}, respectively. } \label{LitFig1}.
\end{figure}

\begin{figure}
\centerline{\includegraphics[width=16 cm]{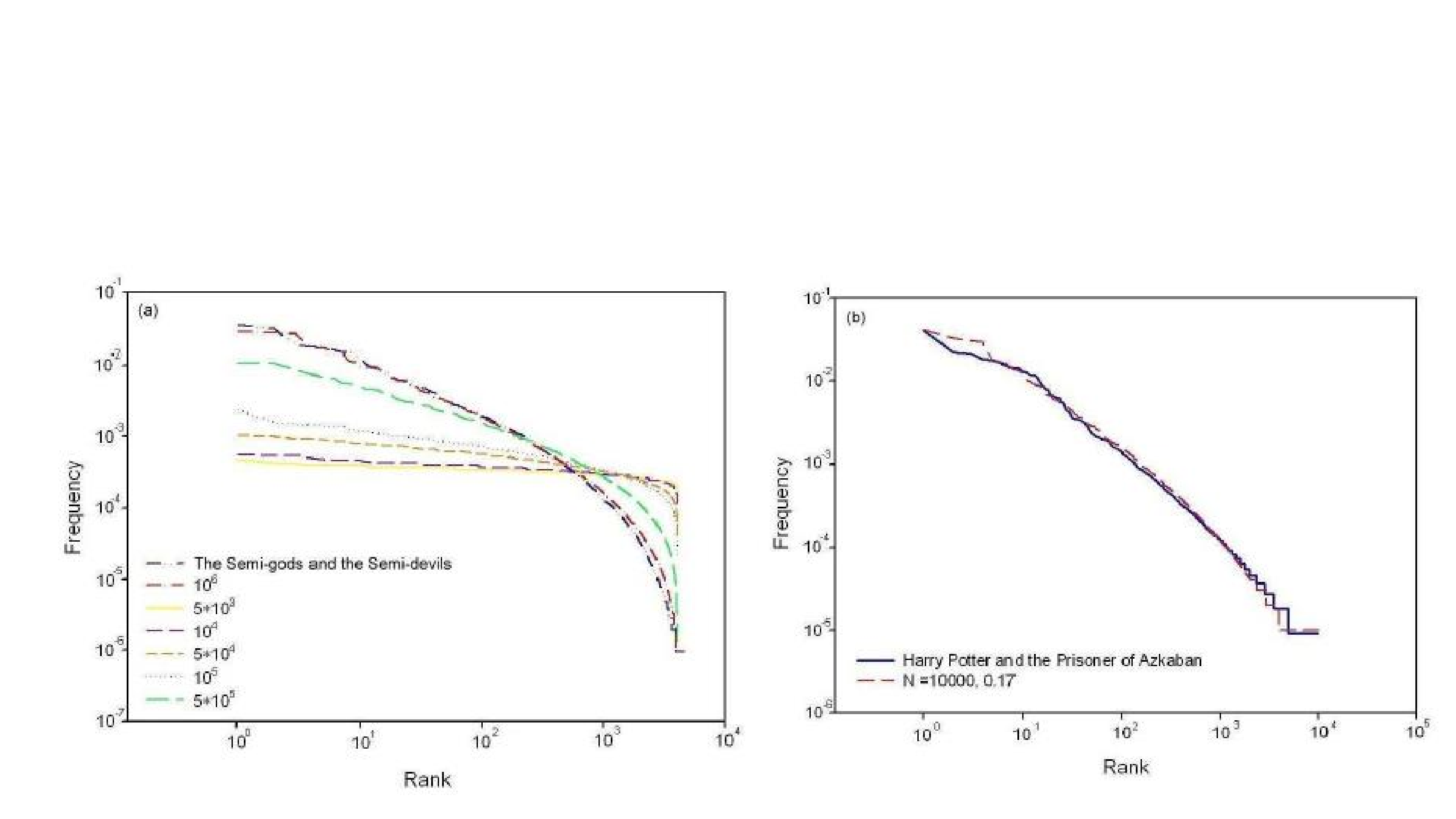}}
\hfill \caption{(Color online) Numerical simulations of Zipf's distributions: (a) number of words $N=$4000,
 exchange rate    $g=0.085$ and simulation steps: $5 \times 10^3$, $10^4$, $5\times 10^4$, $10^5$,  and $10^6$. (b) number of words $N=10000$,
  exchange rate $g = 0.17$ and simulation steps $10^7$. (a) and (b) were taken from Fig. 10(a) and Fig. 10(b) of \cite{05lit},
  respectively.} \label{LitFig2}
\end{figure}

The averaging procedure adopted in HF1MV effectively eliminates
the stretched crossover between the two asymptotic regimes in the
MSLR (dot points and blue line in Fig.~\ref{StockFig1}(b)), so that
the data can be described by the simple  Langevin
equation (Eq.~(\ref{MSDLaggevin})). To facilitate such a
conjecture, we use Eq.~(\ref{MSDLaggevin}) as a guide to write the master equation
\begin{equation}
<r^{*2}>= \tau^*-1+e^{-\tau^*} \label{MSDmaster}
\end{equation}
for fitting  the scaled MSLR $ <r^{*2}>=<r^2>/(D \tau_0)$ versus
$\tau^*=\tau/\tau_0$ for the empirical data.  The data are
well-fitted by $<r^2> \approx \theta \tau^{\alpha_1}$ over the
smaller $\tau$ regime to obtain the pre-factor $\theta =
\frac{D}{2\tau_0}$. We found that the data are well fitted to give
$\alpha_1=2.0$ with an error less than $1\%$. For the larger
$\tau$ regime, the fitting to the asymptotic form $<r^2> \approx D
\tau^{\alpha_2}$ leads to much more scattered values in
$\alpha_2$. To cure this drawback, we impose $\alpha_2=1$ and find
the best fit for $D$.

In Figs.~\ref{StockFig1}(C), we show the scaled curves of
$<r^{*2}>$ versus $\tau^*$, for the 48 months during the years
1996-1999, for 345 stocks from S\&P 500 analyzed in
Fig.~\ref{StockFig1}(b).  The trading hours
for a trading day in 1996-1999 are six and half hours
(9:30AM-4:00PM) for the US market. The time sequences for the collections of
US market are in intervals of 36 seconds. Treating each market
as temporarily stationary over each month, an average $<\cdot>$ is
taken both time-wise and stock-wise. In spite of a slight degree
of bulge at the crossover, we find that the scaled data are in
general in a reasonable agreement with Eq.~(\ref{MSDmaster})
(Fig.~\ref{StockFig1}(C),), that the fitted parameters $\theta$ and $D$
provide useful information about the condition of the market.. In Fig.~\ref{StockFig1}(C),
there are a few curves deviating
significantly from the master curve. These deviations signal
temporarily non stationarity of the markets.

Equation (\ref{MSDmaster}) for the scaled empirical data of Fig. \ref{StockFig1}(c) is very similar to
Eq. (\ref{45cjp}) for the scaled empirical data of Fig. \ref{CPFig1}(c).


\section{Universal patterns in human writing}


In this section, I briefly introduce the results in Section 5 of \cite{05lit} by Hu and Kuo.



In 1949, George Zipf published an interesting result about statistical data in human writing \cite{49Zipf}.
He found that in a given text corpus there is an approximate mathematical relation between the frequency of the occurrence
of each word and its rank in the list of all the words in the text ordered by decreasing frequency \cite{49Zipf}.
The distribution is called the Zipf"s distribution.

\begin{figure}
\centerline{\includegraphics[width=16 cm]{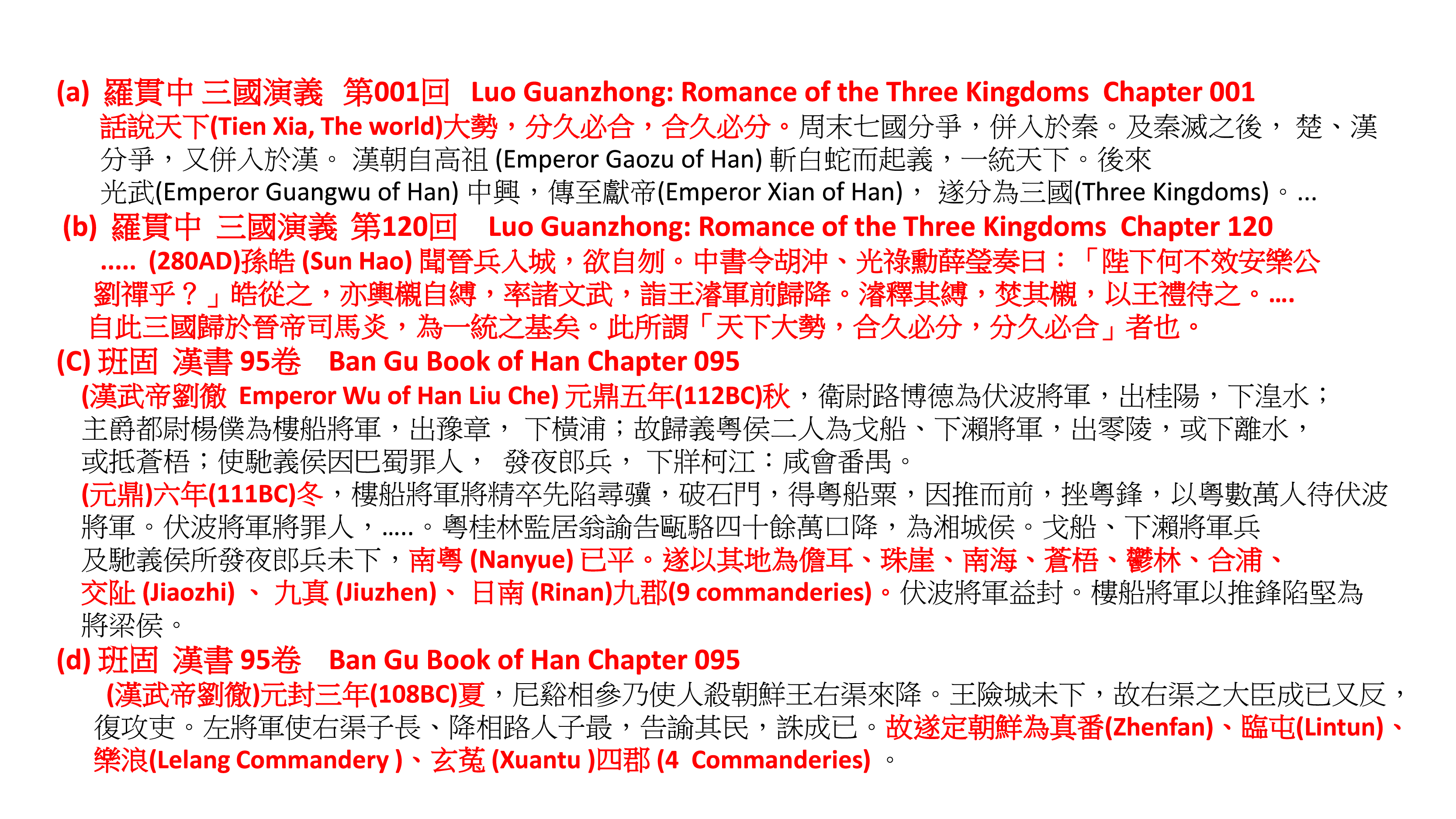}}
\hfill \caption{(Color online) (a) Some statements from Luo Guanzhong: Romance of the Three Kingdoms  Chapter 001
. (b) Some statements from Luo Guanzhong: Romance of the Three Kingdoms  Chapter 120. (c) Ban Gu Book of Han Chapter 095 about Vietnam.
(d) Ban Gu Book of Han Chapter 095
about the Korea.} \label{StateFig1new}
\end{figure}

Zipf's distributions for five Chinese
 novels and five English novels had been calculated and shown in Fig. \ref{LitFig1}(a).
 The Chinese novels were written in Ming Dynasty, Ching Dynasty,
 and recent decades. It is of interest to know that Zipf's distributions for 5 Chinese novels fall approximately
 on the same curve and Zipf's distributions for 5 English novels fall approximately on another curve, which is
 quite different from that of Chinese novels. Thus there are two classes of Zipf's distributions: one for
 Chinese novels and another one for English novels.


It has been found that critical physical systems can be classified into different universality classes so that the systems
in the same class have the same set of critical exponents \cite{71Stanley,2014cjpHuCK}, universal
 finite-size scaling functions \cite{84PF,95prlHu,96prlHu,97preHu,99preA,99preB,03preWHI}, and amplitude ratios \cite{01prlIH}.
  The criteria by which the critical systems can be classified into different universality classes
 is a problem of much academic interest. Figure \ref{LitFig1} suggests Chinese literary works and English literary works could be considered
 as two separate "universality classes". It is of interest to know what are differences between these two classes.
  In the following,  a model proposed by Hu and Kuo \cite{05lit} to generate such curves is introduced.


A writer uses $N$ different words $w_i$, $1 \le i \le N$, in his or her writing and the probability that he
or she uses $w_i$ in the writing is $x_i$,  which satisfy the normalization relation  $\sum_{i=1}^{N} x_i=1$.
We can arrange $x_i$ in a sequence from larger to smaller values and define an integer
 function $I(x_i)$  of $x_i$, which is the order number of $x_i$ in the sequence. We assume that all  $x_i$  are different and
 we can define an inverse function $V $ of $I$  such that $y_I \equiv x_i = V(I(x_i))$.

We assume that a mature writer would have a steady $x_1$, $x_2$, $\dots$, $x_N$ and the corresponding $y_1$, $y_2$, $\dots$, $y_N$,
which was developed following process with an exchange parameter $g$ and a range parameter $r$.

\begin{enumerate}

\item In the beginning, every $x_i$ takes a random number near $1/N$ such that $\sum_{i=1}^{N} x_i=1$. From ($x_1,~ x_2,~ \dots,~ x_N$),
      one has a  corresponding ($y_1,~ y_2,~ \dots,~y_N$) so that $y_i > y_{i+1}$ for $1$  between 1 and $N-1$.

\item Randomly choose an integer $j$ between 1 and $N$, and subtract $g y_j$ from $y_j$:  $y_j \to  y_j - g y_j \ge 0$.

\item Randomly choose an integer $k$ between $j-r$ and $j+r$ and add $g y_j$ to $y_k$ : $y_k \to y_k + g y_j$.

\item Update the order of $y_i$ such that $y_i > y_{i+1}$ for $j$ between 1 and $N-1$.

\item Go to step 2 and iterate the process.

\end{enumerate}

The simulation results for $y_I$ as a function of $I$ in log-log scale for $r=3$, $N = 1000$, $g = 0.1$, and iteration number
 IN = $10^3$, $10^4$, $10^5$, $10^6$ and $10^7$ are presented in Fig. \ref{LitFig1}(b) which
 shows that as IN increases, the Zipf's distributions approach a steady curve similar to the curve for
 Chinese literary works in Fig. \ref{LitFig1}(a).

To understand the differences between curves for Chinese and English novels in Fig. \ref{LitFig1}(a), Hu and Kuo \cite{05lit} tried
to simulate curves in Fig. \ref{LitFig1}(a) and found that with $r$=3, $N$ =4000 and $g$= 0.085 the simulated curves
can approach the curves for Chinese novels as shown in Fig. \ref{LitFig2}(a)
and with $r$=3, $N$ =10000 and $g$ = 0.17, the simulated curves can approach the curve for English novels as shown in Fig. \ref{LitFig2}(b).
Thus the English novels have larger number of different words  $N$ and exchange rate $g$ than Chinese novels.
This is not surprising because the former language is an alphabet spelling system and it is easier to generate new words
to enrich the vocabulary or to use one word to replace another one during development of writing habit.

Zipf's distribution has been used to solve the authorship dispute problem \cite{03PengCK,15ChiTT}.
Shiji prepared by Shima Qian in Western Han Dynasty (202 BC-9 AD) was the most famous historical book in historical China \cite{15ChiTT}
\footnote{In Eq. (2.6) of \cite{15ChiTT}, the sum is over all words $w_i$ in the union of top $R_{cut}$ words in the $i$-th and $j$-th texts.}
The current version of Shiji  contains 130 chapters, including 10 chapters which contains some parts starting with the sentence
"Mr. Chu said ..."; Chang Yen in the Three Kingdoms Period (220-280 AD) considered that such parts were written by Chu Shaosun,
which was widely accepted. We have analyzed the Zipf distributions in Shiji and books written by Liu Xiang (77 BC-6 BC) in Western Han Dynasty.
Our results and other evidences indicate that the parts of Shiji starting
with the sentence "Mr. Chu said ..." might be prepared by Liu Xiang \cite{15ChiTT}. One of the other evidences is Fig. 6 of \cite{15ChiTT}.
The texts in Figs. 6(a) and 6(c) of \cite{15ChiTT} suggest that Chu Shaosun might not be Mr. Chu in Shiji.
After the publication of \cite{15ChiTT}, on 30 May 2017 (Duanwu Festival or Dragon Boat Festival), I learned that Mr. Wang Guowei (1877-1927)
expressed the similar idea in an article ``{\it Han Wei po shi tee ming kao}" which stated that `` Chu Shaosun and Mr. Chu might not be
the same person."

\begin{figure}
\centerline{\includegraphics[width=16 cm]{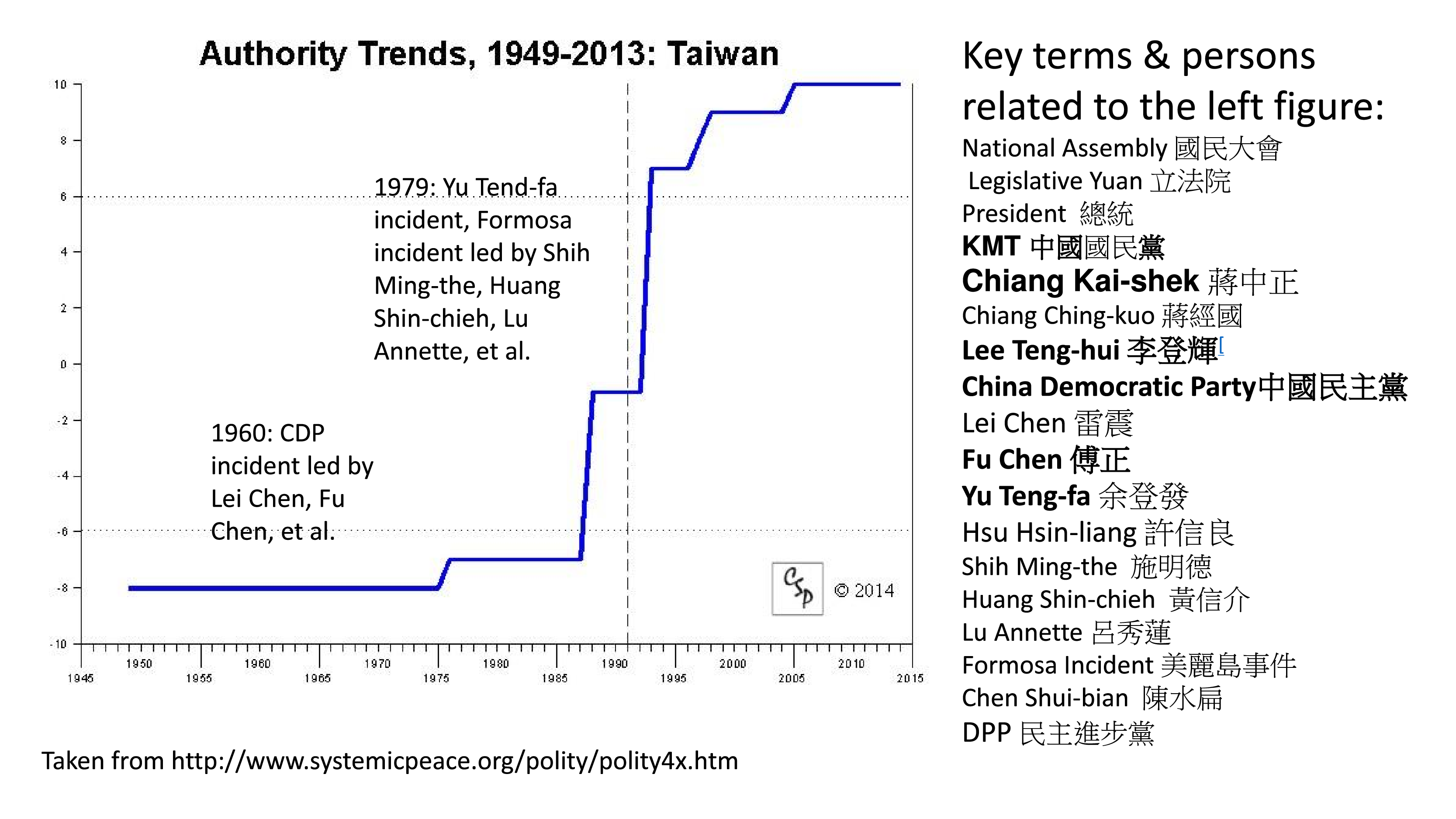}}
\hfill \caption{Chiang Kai-shek was elected as President of the Republic of China on 24 April 1948, 21 March 1954, 21 March 1960, 21 March 1966,
and 21 March 1972.  There was a movement against Chiang Kai-shek's re-election in 1960 led by Lei Chen. After Chiang's re-election in 1960,
 there was a movement to form a new party: China Democratic Party (CDP) Led by Lei Chen. This movement was suppressed by KMT
 successfully by putting Lei Chen in jail.  In 1977-1990, there was another democratic movement. After many
 leaders of this movement were put into jails in early 1979 (Yu Deng-fa) and December 1979 (Formosa incident),
 the movement was continued by family  members of the leaders, lawyers, and students,
and Taiwan became fully democratic. The event in 1960 is similar to a stable point between A and B in Fig. 1(a).
The events in 1979 are similar to point ``b'' (a meta-stable state) in Fig. 1(a). } \label{StatFig2}
\end{figure}

\section{Do countries or states have universal evolution behavior or pattern ?}

Countries or states are composed of a large number of people. Do countries or states have universal behavior
or pattern in their long time evolution?

{\it Romance of the Three Kingdoms} is a historical novel about the Three Kingdoms period in China from
Yellow Turban Rebellion in 184 AD in the Eastern Han Dynasty (23-220 AD) to 280 AD when the Emperor of {\it Eastern Wu}
surrendered to the Jin Dynasty (see below). The novel was first published in the Ming Dynasty (1368-1644AD) and
it was widely believed that the author of the novel was  Luo Guanzhong who was a writer from the end
of the Yuan Dynasty (1271-1368 AD) to the beginning of the Ming Dynasty.

The first chapter of the novel contains a statement about the universal pattern for the long time evolution of states or countries:
``The major trend of the world ({\it Tien Xia} meaning under the heaven) is unification after long period of division,
and division after long period of unification". To support this statement, it was mentioned that after long warring state
period (476-221 BC), Qin established a unified ``Qin Dynasty" (in 221 BC). After the collapse of Qin Dynasty (in 206 BC), Chu and Han competed
with each other and Emperor Gaozu of Han (about 256-195 BC) unified the world (in 202 BC) to establish Han Dynasty....
After long period of unification, Han Dynasty collapsed during Emperor Xian of Han (181-234 AD, as emperor in 189-220 AD)
and enter Three Kingdoms period, see Fig. \ref{StateFig1new}(a).

Three Kingdoms included {\it Cao Wei} (220-266 AD) in the Yellow River region,  {\it Shu-Han} (221-263 AD) in Sichuan, and
{\it Eastern Wu} (222-280 AD) in the Yangtze River region, the Pearl River region, and North Vietnam.
Cao Cao (155-220 AD) established the foundation for {\it Cao Wei} and controlled Emperor Xian of Han in 196-220 AD.
In 220 AD, Cao Pi, Cao Cao's son, forced Emperor Xian of Han to step down and established {\it Cao Wei}.
{\it Shu-Han} was established by Liu Bei (161-223 AD, as emperor in 221-223 AD).
In 263AD, Liu Shan (207-271 AD, as emperor in 223-263 AD), Liu Bei's son, surrendered to {\it Cao Wei}.
In 265 AD, Cao Huan (246-302 AD, as emperor in 260-265 AD),
the emperor of {\it Cao Wei}, was forced to step down by Sima Yan (236-290 AD, as emperor in 266-290 AD) who established Jin Dynasty
and was known as Emperor Wu of Jin. {\it Eastern Wu} was established by Sun Quan (182-252 AD, as emperor in 222-252 AD).
In 280 AD, Sun Hao (243-284 AD, as emperor in 264-280 AD), the emperor of {\it Eastern Wu} and the grandson of Sun Quan, surrendered
to Sima Yan and ended the Three Kingdoms period; at the end of the last chapter of {\it Romance of the Three Kingdoms},
there was a comment on this event:``The major trend of the world ({\it Tien Xia}) is
division after long period of unification and unification after long period of division", see  Fig. \ref{StateFig1new}(b).

It seems that what Luo Guanzhong had in mind about {\it Tien Xia} included only the regions of the Yellow River
and the Yangtze River; he wrote that Emperor Gaozu of Han unified the world as shown in Fig. \ref{StateFig1new}(a),
but Emperor Gaozu of Han never took Nanyue in the Pearl River region and North Vietnam into the territory of the Han Dynasty
while Nanyue belonged to the territory of Qin Dynasty (221-206BC). Zhao Tuo (about 240 BC to 137 BC) was a general of Qin Dynasty.
After the collapse of the Qin Dynasty, Zhao Tuo  established Nanyue as an independent state with the capital at Panyu (now Guangzhou).

Book of Han by Ban Gu is a book about Han Dynasty. Figures \ref{StateFig1new}(c) and \ref{StateFig1new}(d) contain
the historical records that Emperor Wu of Han took Nanyue (including Jiaozhi, Jiuzhen, and Rinan in the middle and north Vietnam)
and Korea (including Zhenfan, Lintun, Lelang and Xuantu 4 Commanderies)
into the territory of Han Dynasty in 111BC and 108 BC, respectively. The text in Figures \ref{StateFig1new}(b) indicated
that Luo Guanzhong considered that Emperor Wu of Jin had unified {\it Tien Xia} in 280AD, but in this year the original
territories of Zhenfan, Lintun, Lelang and Xuantu did not belong to Jin Dynasty. During the life time of Luo Guanzhong,
the original territories of Jiaozhi, Jiuzhen, and Rinan did not belong to Yuan or Ming Dynasties.
Thus  Luo Guanzhong's statements about {\it Tien Xia} in Figs. \ref{StateFig1new}(a) and \ref{StateFig1new}(b)
could be supported by historical records when {\it Tien Xia} included only the regions of the Yellow River and the Yangtze River.

Dr. Monty G. Marshall, a Principal Investigator of  Societal-Systems Research Inc., has executed a
Polity IV Project with the website: http://www.systemicpeace.org/polity/polity4.htm, which
contains data about Annual Polity scores (APS) from -10 to 10 for 167 countries in 1946-2013.
The countries include following categories with related scores: Full democracy (10), Democracy (6 to 9),
Open Anocracy (1 to 5), Closed Anocracy (-5 to 0), Autocracy (-10 to -6), Failed or Occupied.
From the data in this website, one can easily find that from 1980 to 1990 to 2010, Annual Polity scores of most countries
increase or maintain at the highest value 10. Figure in the web-site
http://www.systemicpeace.org/polity/polity1.htm shows that the number of democracy countries increases and
the number of autocracy countries decreases after 1980.

The Annual Polity scores of Taiwan can be found in Fig. \ref{StatFig2}.
Here I would like to explain the data in  Fig. \ref{StatFig2} and present a simple model for the evolution pattern.

According to the Republic of China (ROC) Constitution passed by National Assembly on 25 December 1946 in Nanjing
and executed from 25 December 1947,
National Assembly is responsible for election of ROC President and Vice President, and the modification of the Constitution.
Executive Yuan is responsible for administration of ROC Central Government.
Legislative Yuan is responsible for passing laws and government budgets. Members of National Assembly and Legislative Yuan
should be re-elected every 6 and 3 years, respectively.
A person can be ROC President for consecutive two terms and each term has 6 years.
Chiang Kai-shek (31 October 1887-5 April 1975) was elected as ROC President on 20 April 1948 in Nanjing.

KMT (Kuomingtang, China National Party) failed in the war with China Communist Party in China and moved ROC Central Government,
National Assembly, Legislative Yuan, etc, from Nanjing to Taipei in 1949. Thus the data
in Fig. \ref{StatFig2} started in 1949. Since members of the National Assembly and Legislative Yuan were moved from China to Taiwan
and they were not re-elected,  Taiwan Annual Polity scores from 1949 to 1974 were at very low level of -8.

The imposition of Taiwan martial law started on 20 May 1949, and the publication of new newspapers and the formation of new political
parties were prohibited. When Wu Kuo-Cheng (1903-1984) was Taiwan governor in 1949-1953, Taiwan started elections of city mayors,
county heads, and councillors via popular votes in 1950-1951. Most elected persons were KMT members, but there were still some non KMT members,
e.g.  Wu San-Lien (1899-1988) was elected as Taipei mayor on 7 January 1951, Henry Kao (1913-2005) was elected as Taipei mayor on 2 May 1954
and Yu Deng-fa (1904-1989) was elected as Kaohsiung county head on 24 April 1960.

Chiang Kai-shek was elected as ROC President on 21 March 1954, 21 March 1960, 21 March 1966, and 21 March 1972 in Taipei
by the National Assembly. In the 1966 and 1972 elections, Yen Chia-kan (1905-1993) was elected as the Vice President.
In 1960, National Assembly managed to modify the ROC Constitution so that Chiang Kai-shek could be elected as ROC President
for the third term, the 4th. term, etc.
There was a movement against the modification of the ROC constitution and Chiang Kai-shek’s re-election in 1960 led by Lei Chen (1897-1979)
who was the key responsible person of the periodical Free China Journal (1949-4 September 1960).
After the ROC constitution was modified and Chiang Kai-shek's election into the third term in 1960, there was a movement
to form a new party:  China Democratic Party (CDP) led by Lei Chen. This movement was suppressed by KMT by
 putting Lei Chen and his supporter Fu Chen (1927-1991) in jail on 4 September 1960. Lei Chen was in jail for 10 years and
Fu Chen lost his freedom for 6 years. Free China Journal was also terminated. Other participants of the democratic
movement in 1960 did not continue the movement and KMT thus successfully suppressed
the democratic movement in 1960.

From 1969, Taiwan citizens began to elect a small
number of persons to join the original large numbers of members in National Assembly and Legislative Yuan,
e.g. Huang Hsin-chieh (1928-1999) and Kang Ning-hsiang (born 1938) were first elected as members of Legislative Yuan in 1969 and 1972,
respectively, and thus Annual Polity scores had slightly increased from 1976 to 1986. There were bigger jumps after 1986.
Such bigger jumps could be related to democratic movement after 1977.

In 1971, ROC government in Taipei lost the seat in United Nations (UN), more and more people,
including students, paid attention to the revision to the political system.
In May 1972, Chiang Ching-kuo (27 April 1910-13 January 1988) was appointed by his father Chiang Kai-shek as the President
of the Executive Yuan. Chiang Ching-kuo paid more attention to economic
development in Taiwan and had a stronger control over Taiwan society. On 23 December 1972, all elected mayors and county heads
were KMT members, Hsu Hsin-liang (born 27 May 1941), also a KMT member, was elected as a councillor of Taiwan Provincial Council.

 On 5 April 1975,  Chiang Kai-shek passed away and Yen Chia-kan became the ROC President; punishments for all prisoners were reduced and
Shih Ming-teh (born 15 January 1941, as a political prisoner since 1962) was released on 16 June 1977. On 19 November 1977,
Taiwan had election of city mayors, county heads, and councillors. Huang Hsin-chieh and Kang Ning-hsiang organized
collective champion activities to support non-KMT candidates, called Tangwai, meaning outside KMT. Hsu Hsin-liang was not nominated
by KMT and run for Taoyuan county head as an independent candidate. A voter in Zhongli city of Taoyuan county reported
to witness the KMT rigging the election, which leaded
to Zhongli incident, that made electoral fraud greatly reduced in all Taiwan in this election. Among 20 seats of mayors and county heads,
KMT lost 4 seats including Taoyuan county; among 77 seats of councillors of Taiwan Provincial Council, KMT lost 21 positions. The later
included Lin Yi-hsiung (24 August 1941) and Chang Chun-hung (born 17 May 1938).

On 20 May 1978, Chiang Ching-kuo took the position of ROC President.
On 23 December 1978, there were planed elections of additional members for National Assembly and Legislative Yuan. The election was canceled
due to the announced US-PRC diplomatic relation starting on 1 January 1979. In January 1979,  Yu Deng-fa invited Tangwai to his home town for a meeting
on 1 February 1979 to exchange ideas about what to do in the next step. Before the meeting, Yu Deng-fa and his son were arrested.
Hsu Hsin-liang, Shih Ming-teh, et al. organized a protest demonstration in Yu Deng-fa's home town Qiaotou. This Qiaotou incident lead KMT
to freeze Hsu Hsin-liang's position of Taoyuan county head for two years, Hsu then went to USA.

Huang Hsin-chieh, Shih Ming-teh, Lin Yi-hsiung, Chang Chun-hung, Yao Chia-wen (born 15 June 1938), Lu Annette (born 6 June 1944),
Chen Chu (born 10 June 1950), Lin Hung-hsuan (1942-2015), et al. continued the democratic movement and published the political magazine ``Formosa".
Some leaders of the magazine organized a march  in Kaohsiung to celebrate the world human-right day on 10 December 1979.
The activity was constrained by policemen and military personnel,
which caused the conflict between two sides and leaded to Formosa Incident, also known as Kaohsiung Incident \cite{791210}. Most active members of
the Formosa magazine were arrested from 13 December 1979 and Formosa magazine was terminated. Shih Ming-teh escaped with the help
from the Presbyterian Church and was arrested on 8 January 2000.
Lawyers, e.g. Chang Teh-ming (born 1938), Chen Shui-bian (born 12 October 1950), Frank Hsieh (19 May 1946), Su Tseng-chang (28 July 1947), et al.
helped arrested persons in the court.  Shih Ming-teh, Huang Hsin-chieh, Lin Yi-hsiung, Chang Chun-hung, Yao Chia-wen,
Lu Annette, Chen Chu, and Lin Hung-hsuan were accused in the military course for overthrowing the government.
On 28 February 1980, the first day of the court process, Lin Yi-hsiung's mother and twin daughters were killed to
death in the basement of Lin's home. Lin's elder daughter was also killed, but found earlier and sent to a hospital to rescue,
and is still alive. On 18 April 1980, Shih Ming-teh and Huang Hsin-chieh were sentenced to life in prison and 14 years in prison, respectively;
other 6 persons were sentenced to 12 years in prison \cite{791210}. In April-May 1980, another group of 33 people were tried in civil court
and sentenced to terms ranging from two to six years \cite{791210}. The third group of persons  helped Shih Ming-teh to escape;
this group included Kao Chun-ming (born 6 June 1929) and Chang Wen-ing (born 26 July 1950) who were sentenced to 7 and 4 years in prison, respectively.

On 6 December 1980, the supplementary elections for the National Assembly and the Legislative Yuan canceled in December 1978 were restored.
Among 70 directly elected members of Legislative Yuan, Tangwai took 9 seats, including Kang Ning-hsiang, Chang Teh-ming,
Huang Tien-fu (born 1938, Huang Huang Hsin-chieh's younger brother) and Hsu Jung-shu (27 December 1939, Chang Chun-hung's wife).
Among 76 elected members of National Assembly, Tangwai took 14 seats, including Chou Ching-yu (born 12 June 1944, Yao Chia-wen's wife)
who got the largest number of votes in all Taiwan. Thus Tangwai had slightly recovered from the great damage of Formosa Incident \cite{791210}.

On 20 May 1984, Chiang Ching-kuo took the position of ROC President for the second term, and Lee Teng-hui
(born 15 January 1923) \cite{LeeTH} took the position of ROC Vice President.

On 15 October 1984, Henry Liu (1932-15 October 1984) was murdered in the garage of his home in Daly City, California \cite{19841015}.
It was found that ROC Military Intelligence Bureau, Vice Admiral Wang Hsi-ling ordered Chen Chi-li, a gangster head, to do so.
This event greatly damaged the image of KMT government.

On 28 September 1986, Tangwai announced to establish a new political party: Democratic Progressive Party (DDP) in a meeting in the Grand Hotel.
This time KMT did not try to suppress Tangwai as in 1979. On 7 October 1986, Chiang Ching-kuo told Katherine C. Graham of Washington Post that
his government would release the martial law, allow the publication of new newspapers, and formation of new political parties.
In the election on 6 December 1986, among 55 regional seats of elected new members
of the Legislative Yuan, DPP took 11 seats; among 84 seats of the National Assembly, DDP took about 15 seats.

On 15 July 1987, Taiwan martial law was lifted. On 13 January 1988, Chiang Ching-kuo passed away, and
Lee Teng-hui became ROC (Taiwan) president.
In March 1990, National Assembly had a meeting in Taipei to elect ROC president and Vice President.
There were 752 members and only 84 persons were elected in Taiwan on 6 December 1986, and other members were elected before 1949.
Such a situation inspired a student movement, called  Wild-Lily student movement \cite{90student} starting on 16 March 1990
at the  Chiang Kai-Shek (CKS) Memorial Hall.
On 18 March 2018, the representatives of students made their four demands \cite{90student}:
1. Abolish the National Assembly and re-establish a new National Assembly infrastructure,
2. Nullify the Temporary Provisions Against the Communist Rebellion and re-establish constitutional order,
3. Hold a National Affairs Conference,
4. Establish a political reform time table.

On 20 March 1990, there were more than 5,000 protestors.
On 21 March 1990, Lee Teng-hui was elected as ROC president. After the election, President Lee invites 50 students to the
Presidential Hall and gave positive response to the students' request. The students  ended the movement on 22 March 1990.
In May 1990, President Lee announced the invalidation of trials for the Formosa Incident \cite{791210}.
In June 1990, Justices of Constitutional Court of the Judicial Yuan made Explanation 261:
All members of National Assembly and Legislative Yuan elected before 1949 should step down before 31 December 1991.

On 21 December 1991, there were elections of 325 members of the Second National Assembly.  In this election, KMT took 254 seats, DDP took 66 seats,
other parties and independent persons took 5 seats. All members of National Assembly and Legislative Yuan elected before 1949
had stepped down before 31 December 1991. From 1 January 1992,  National Assembly and Legislative Yuan operated by representatives
elected in Taiwan and Annual Polity Scores of Taiwan in Fig \ref{StatFig2} had reached 7 and Taiwan became a democratic country.
On 19 December 1992, there were elections of 161 members of the Second Legislative Yuan, in this election KMT took 95 seats, DDP took 51 seats,
other parties and independent candidates took 15 seats.

In July 1994, the Second National Assembly revised the ROC Constitution, so that ROC President and Vice President
should be elected directly by citizens in Taiwan region via a popular vote.
On 23 March 1996, Taiwan citizens elected Lee Teng-hui as ROC (Taiwan) President and Lien Chan (born 27 August 1936) as ROC (Taiwan)
Vice President.

On 18 March 2000, Taiwan citizens elected DDP members Chen Shui-bian as ROC (Taiwan) President and Lu Annette as ROC (Taiwan) Vice President.
This was the first time that DDP won the presidential election, and there was a peaceful transfer of power from one party to another;
the Annual Polity Score of Taiwan in Fig \ref{StatFig2} had reached 9.
On 20 March 2004, Taiwan citizens again elected Chen Shui-bian as ROC (Taiwan) President and Lu Annette as ROC (Taiwan) Vice President
 and the Annual Polity Score of Taiwan in Fig \ref{StatFig2} had reached the highest value 10. From 20 May 2000 to 20 May 2008, DDP took the
 President Office, but the number of seats DDP had in the Legislative Yuan was smaller than 50\%. From 20 May 2008 to 20 May 2016,
 KMT took the President Office and controlled the Legislative Yuan. On 16 January 2016, DDP won both President election
 and election of the Legislative Yuan. From 20 May 2016, DDP took the President Office and controlled the Legislative Yuan and thus
 could passed some laws so that political parties in Taiwan can compete on more equal basis. For the election on 16 January 2016, KMT donated
 KMT president candidate more than $2 \times 10^8$ new Taiwan dollars, and DDP donated DDP president candidate less than $10^7$ new Taiwan dollars.

Here I would like to present a simple model for the transition of Taiwan from an autocracy country in 1949-1975 to a full democracy
country after 2004 as shown in Fig \ref{StatFig2} by using the analogy of such a transition with behavior of the gas-liquid
system shown in Figure \ref{CPFig1}(a).

Figure \ref{CPFig1}(a) is a phase diagram for gas-liquid system of CO$_2$. A$^{\prime}$CE$^{\prime}$ curve is an isothermal curve passing
the critical point C with the critical temperature $T_c=31.3^\circ$C. A-B-C-D-E is an isothermal curve for the temperature $T=21.5^\circ$C.
below the critical temperature $T_c$.  A to B is in the liquid phase.  When there is a fluctuation, e.g. due to cosmic ray radiation,
  at the point between A and B, the fluctuation can be suppressed easily.
 When there is no impurity in the system and the volume increases, the system can maintain at the liquid state and reach point b,
 which is a meta-stable state. When there is a fluctuation, the system at point b can decay
 easily and reach a stable state between B and D, which is a mixture of liquid and gas.

 In 1949-1975, Taiwan was an autocracy country and was similar to liquid state
 in Fig. \ref{CPFig1}(a). The democratic movement lead by Lei Chen in 1960 was similar to  a fluctuation
 at a point between A and B in Fig. \ref{CPFig1}(a)
 and it was suppressed by KMT easily. Due to economic growth, more and more people received higher education, etc, but KMT still controlled
 the society tightly via Taiwan martial law, Taiwan reached a meta-stable state in 1976-1985, similar to point b
 in Fig. \ref{CPFig1}(a); Yu Teng-fa incident and Formosa incident \cite{791210}
 in 1979 shown in Fig \ref{StatFig2} were similar to fluctuations near point b in Fig. \ref{CPFig1}(a). KMT could not suppress such
 fluctuations and Taiwan was approaching to a two-parties political system after 1986, similar to a point in the horizontal line BC in
 Fig. \ref{CPFig1}(a), which is a mixture state.

 It seems that the model for the increase of annual polity scores of Taiwan can be
 applied to other countries, thus the number of democracy countries increases and
the number of autocracy countries decreases after 1980 as shown in http://www.systemicpeace.org/polity/polity1.htm.
The time scale for such increase might depend on the system size. For a large country, it might take longer time
to transfer from an autocracy country to a democracy country.



\section{Summary and discussion}




We have shown that
concepts and methods of statistical physics could be useful for understanding universal
and scaling behaviors in complex many-body systems, including human and social systems.
We are using such concepts and methods to study evolution of root networks of Chinese characters \cite{root}
and universal behaviors and patterns in Chinese history.

Results of Section 3 inspire other problems for further studied as discussed in \cite{05lit}.

There exists other democracy index, e.g. https://en.wikipedia.org/wiki/Democracy\_Index, but data for such index do not
cover many years  as Fig. \ref{StatFig2} for studying evolution pattern.

\section{Acknowledgments}

The author thanks collaborators, R. E. Amritkar,   Wen-Jong Ma, Shih-Chieh  Wang,  Chi-Ning  Chen,
Wei-Cheng Kuo, for the work reported in paper. He thanks Armen Allahverdyan, Monty G. Marshall, Johannes Voit,
and Karen Petrosyan for critical reading of the paper.
He also thanks Deng Weibing and Wei Li for inviting him to give a talk at
The 5th International Workshop on Statistical Physics and Mathematics for Complex Systems (SPMCS2017),
in Wuhan, China on 11-15 October 2017; this paper was based on this talk.
This work was supported by Grant MOST 106-2112-M-001-027.

\end{document}